\documentclass[pre,preprint]{revtex4-1}

\usepackage{graphicx}
\usepackage{amssymb}
\usepackage{amsmath}
\usepackage{amsthm}
\usepackage{epstopdf}

\begin{document}

\title{Microcanonical Monte Carlo Study of One Dimensional Self-Gravitating Lattice Gas Models}

\author{J.~M.~Maciel}
\email{marco@fis.unb.br}
\author{M.~A.~Amato}
\author{T.~M.~Rocha Filho}
\author{A.~D.~Figueiredo}
\affiliation{Instituto de F\'{i}sica and International Center for Condensed Matter Physics, Universidade de Bras\'\i lia, CP 04455, 70919-970 - Bras\'\i lia, Brazil }


\begin{abstract}
In this study we present a Microcanonical Monte Carlo investigation of one dimensional self-gravitating toy models. We study the effect 
of hard-core potentials and compare to those results obtained with softening parameters and also the effect of the geometry of the models.
In order to study the effect of the geometry and the borders in the system we introduce a model with the symmetry of motion in
 a line instead of a circle, which we denominate as $1/r$ model. 
The hard-core particle potential introduces the effect of the size of particles and, consequently, the effect of the density of the 
system that is redefined in terms of the packing fraction of the system. The latter plays a role similar to
the softening parameter $\epsilon$ in the softened particles' case. In the case of low packing fractions both models with hard-core
 particles show a behavior that keeps the intrinsic 
properties of the three dimensional gravitational systems such as negative heat capacity.
For higher values of the packing fraction the ring
the system  behaves as the Hamiltonian Mean Field  model and while for the $1/r$ it is similar to the one-dimensional systems.
\end{abstract}

\maketitle

\section{Introduction}
\label{sec1}

Many particles systems with long-range interactions present a plethora of novel and challenging problems in comparison to
short-range interacting systems: non-Gaussian quasi-stationary states, ensemble non-equivalence 
and the existence of regions of negative heat capacity~\cite{campa,campa2}. Examples of such systems include non-neutral plasmas~\cite{rizzato},
self-gravitating systems~\cite{tanu}, vortices in two-dimensional turbulent hydrodynamics~\cite{venaille},
free electron laser~\cite{anton1}, and some models of interest in the literature as one and two-dimensional self-gravitating systems~\cite{mila,teles},
the Hamiltonian Mean Field (HMF) and ring models~\cite{antoni,sota}.
A long-range interaction potential decays at large distance as $r^{-\alpha}$ with $\alpha \leq d $, where $d$ is the spatial dimension
implying that the total potential energy increases superlinearly with the volume~\cite{campa,dauxois}. 

For a stationary state of a many-body isolated self-gravitating system, and therefore described by the microcanonical ensemble, in a three-dimensional space, the virial theorem states that
$E = K+V=-K$,  with $K$ and $V$ the total kinetic and potential energies, respectively, which results in negative heat capacity.
However it is well known that in the canonical ensemble the heat capacity is always positive, therefore the measures in both ensembles are inequivalent.
This fact can be restated by noting that in this case the Legendre transformation of the microcanonical internal energy
is not invertible, and the entropy being non-additive the usual Maxwell construction is not applicable. This implies that
stable states in the microcanonical ensemble can be metastable or unstable in the canonical ensemble~\cite{ruffo}.
In this sense, the microcanonical ensemble is the most fundamental one and conveys all the required information on
the thermodynamical properties of long-range interacting systems~\cite{campa,tanu,lynden}.

The main difficulty in the study of three-dimensional self-gravitating systems is the short distance divergence of the potential and
particle evaporation, meaning that the entropy has local maximum instead of a global maximum.
Thus, no real stationary state is possible unless the system is
enclosed in a finite volume. Even in this case, if the short-range divergence is not regularized, it may lead to the phenomenon of gravothermal catastrophe\cite{tanu,lynden,antonov}.
A common regularization consists to introduce a small softening parameter $\epsilon$ in the form $V(r)\propto 1/(r+\epsilon)$~\cite{heggie}.
In order to avoid these difficulties, lower dimensional models have been introduced to study the gravitational systems by considering additional symmetries, as
for one and two-dimensional self-gravitating systems and the ring model. Such models have played an important role in the study of properties
of long-range interacting systems for the simplifications they allow to introduce in the performed calculations. In the present paper we intend to show that a further simplification level is possible by
introducing the lattice-gas counterpart of such models, where a drastic simplification of the microscopic state is obtained by considering a local average
of the exact $N$-body dynamics. Here we restrict ourselves to the determination equilibrium properties from Monte Carlo simulations using
a lattice of points as configuration space, with finite spacing $l$,
in such a way that the results converge to those obtained from the exact microscopic dynamics.
We are also interested in studying the effects of a hard-core contribution of the interparticle potentials and compare it to using a softening regularization parameter,
and as well the physical consequences of different boundary conditions.

The paper is structured as follows: in Section~\ref{sec2} the models considered here are introduced.
In Section~\ref{sec3} a discussion of the numerical methods is provided,
and in Sections~\ref{sec4} and~\ref{sec5} we present the results for the two models considered here. The limitations of the Kac prescription
for potential with a hard-core part is discussed in Section~\ref{sec6}, followed some concluding remarks in Section~\ref{sec7}.


\section{Models}
\label{sec2}

We briefly introduce the continuous models studied in the remaining of the paper and their respective
lattice-gas description. All models have a Hamiltonian of the form
\begin{equation}
H= \sum_{i=1}^{N} \frac{p_i^2}{2m} + \frac{1}{N} \sum_{i<j}^{N} V(x_i-x_j),
\label{hamil_sgl}
\end{equation}
where the interparticle potential $V(x)$ is specified below for each model.
Throughout the paper the mass of all particles are taken to be unity.
The self-gravitating ring model was introduced by Sota et al.~\cite{sota}, and consists of $N$ point particles constrained on a ring of fixed radius and interacting by three-dimensional Newtonian gravity. 
Here we consider two different ways to avoid the divergence of the potential at zero distance.
In the first case a hard-core part is introduced at short distances in the potential:
\begin{equation}
V(\theta_i-\theta_j)= \left \{ \begin{array}{lc}
-1/\sqrt{2(1- \cos(\theta_i - \theta_j))} & \mbox {, if } \theta_i - \theta_j\geq l; \\               
+ \infty & \mbox{, if } \theta_i - \theta_j < l,
\end{array} \right .
\label{sgr_hard}
\end{equation}
and  $l$ is a fixed positive constant (size of the particle) and the coordinates $x_i$ are identified with the position angles $\theta_i$.
The other regularization method consists in introducing a small softening parameter $\epsilon$ such that:
\begin{equation}
V(\theta_i-\theta_j)= \frac{-1}{\sqrt{2} \sqrt{1- \cos(\theta_i - \theta_j)+\epsilon}},
\label{sgr_smooth}
\end{equation}
which is the original form considered in Ref.~\cite{sota}.
Taking the large $\epsilon$ limit in Eq.~(\ref{sgr_smooth}) we obtain
\begin{equation}
\label{sgr_li}
  V(\theta_i, \theta_j)= \frac{1}{\sqrt{2 \epsilon}} {\Bigg[} \frac{1-\cos(\theta_i - \theta_j)}{2 \epsilon} -1 {\Bigg]} + \mbox {{\it O}}(\epsilon^{-2}),
\end{equation}	
which is the potential for the Hamiltonian Mean-Field (HMF) model,
a paradigmatic model widely studied in the literature~\cite{antoni,campa,campa2}.
The HMF model has two phases, a low energy phase with particles in a single 
cluster and a high energy gas phase with particles homogeneously distributed on the circle,
with a second order phase transition connecting the two phases~\cite{antoni}.

The order parameter (magnetization) for both the ring and HMF models is defined as:
\begin{equation}
M = \sqrt{{M_x}^2 + {M_y}^2 },
\label{magnetization_sgr}
\end{equation}
with the magnetization components:
\begin{equation}
M_x=\frac{1}{N}\sum_{i=1}^N \cos(\theta_i),\hspace{5mm}
M_y=\frac{1}{N}\sum_{i=1}^N \sin(\theta_i).
\label{magcomps}
\end{equation}

Both previous models have periodic boundary conditions. So in order to study the effects of the (non-periodic) borders on the thermodynamic properties,
we introduce a one-dimensional toy model of $N$ particles in a line of finite length $L$, with a $1/r$ type of potential:
\begin{equation}
V(x_i-x_j)=\left\{
\begin{array}{lc}
-1/|x_i - x_j|& \mbox {, if } |x_i - x_j| \geq l; \\
+\infty & \mbox{, if } |x_i - x_j| < l,
\end{array}
\right . 
\label{sgl_hard}
\end{equation}
with a hard-core regularization, and
\begin{equation}
V(x_i, x_j)= -\frac{1}{|x_i - x_j|+ \epsilon},
\label{sgl_smooth}
\end{equation}
with a regularizing softening parameter $\epsilon$.
In the large $\epsilon$ limit, the potential in Eq.~(\ref{sgl_smooth}) becomes
\begin{equation}
\label{sgl_li}
  V(x_i, x_j)= \frac{1}{\epsilon^2} (|x_i - x_j|- \epsilon ) + \mbox {{\it O}}(\epsilon^{-3}),
\end{equation}
which is the potential of the one-dimensional self gravitating model, formed by infinite sheets
of finite mass moving only in the normal direction to the planes ~\cite{hohl}. Its heat capacity is always positive.


\section{Microcanonical Monte Carlo Algorithm}
\label{sec3}

Computer simulations using Monte Carlo methods are extensively used in theoretical physics \cite{allen,kalos,landau,binder,maciel}.
In its original form the system configurations is determined using the Metropolis algorithm~\cite{metropolis}, which
generates the equilibrium configurations in the canonical ensemble. Due to the possibility of ensemble inequivalence
for systems with long-range interactions, the choice of the working ensemble, as the most fundamental,
is the microcanonical one, in the sense that it delivers all physical information of the equilibrium state.

Probably the most popular implementation of a microcanonical Monte Carlo method is due to Creutz~\cite{creutz}, where an extra degree of freedom called a demon
performs a random walk on different configurations of the system, with the total energy of both the system and of the demon preserved. If the system
is sufficiently large then one can consider its energy as fixed,
with only little fluctuations due to the demon. For small systems this may represent
a limitation of the algorithm, and since small long-range interacting systems, with
just a few hundred particles, are of some interest,
we prefer here the approach developed by Ray~\cite{ray}, with the energy being exactly conserved, even for very small systems.

The (volume) microcanonical probability density can be written as
\begin{equation}
\Omega_E(E-H(\{p_i,q_i\}))=C(N)\:\Theta(E-H(\{p_i,q_i\})),
\label{mcpd}
\end{equation}
where $\Theta(X)$ is the Heaviside function and $C(N)$ a normalization constant. By integrating $\Omega_E(E-H(\{p_i,q_i\}))$ over the momenta,
we obtain the probability density in configuration space~\cite{ray}:
\begin{eqnarray}
W_E(\{ q_i\}) & = & \int {\rm d}{\bf p}_1\cdots {\bf p}_N\:\Omega_E(E-H(\{p_i,q_i\}))
\nonumber\\
 & = & \overline{C}(N) [E- V (\{ q_i\}) ]^{\frac{dN}{2} -1}\:\Theta [E- V (\{ q_i\}) ],
\label{density}
\end{eqnarray}
where $\overline C(N)$ is a constant depending only on the number of particles $N$. The acceptance probability of a Monte Carlo move
$\{q_i\}\rightarrow\{q'_i\}$ is thus given by:
\begin{equation}
P_E (\{ q_i\} \rightarrow \{ {q'}_i\})= \min \left( 1  ,  \frac{[K (\{ q'_i\}) ]^{\frac{dN}{2} -1} \Theta [K (\{ q'_i\}) ]}
{[K (\{ q_i\}) ]^{\frac{dN}{2} -1}} \right),
\label{acceptance}
\end{equation}
with $K(\{ q_i\})= E - V(\{ q_i\})$ the kinetic energy.

For a lattice gas, the configuration space is composed by a set of equally spaced lattice points.
If the inter-particle potential has a hard-core, then only one particle can occupy each site of the lattice,
and the size or the hard-core is here considered to be given by the lattice spacing $l$.
The continuous limit is approached when the number of lattice points $N_{l}$ becomes very large.

\section{Self-Gravitating Ring Model}
\label{sec4}

In order to bring forward the scaling of the potential energy with $N$
in the simulations, it is rescaled as $V\rightarrow V/|V_{\rm min}|$ with $V_{\rm min}$ being the absolute minimum of the potential.
In this way the minimum energy per particle for all models is $e_{\rm min}=-1$.

\subsection{Softening parameter regularization}
\label{sec4a}

\begin{figure}[htbp]
\centering
\scalebox{0.27}{\includegraphics{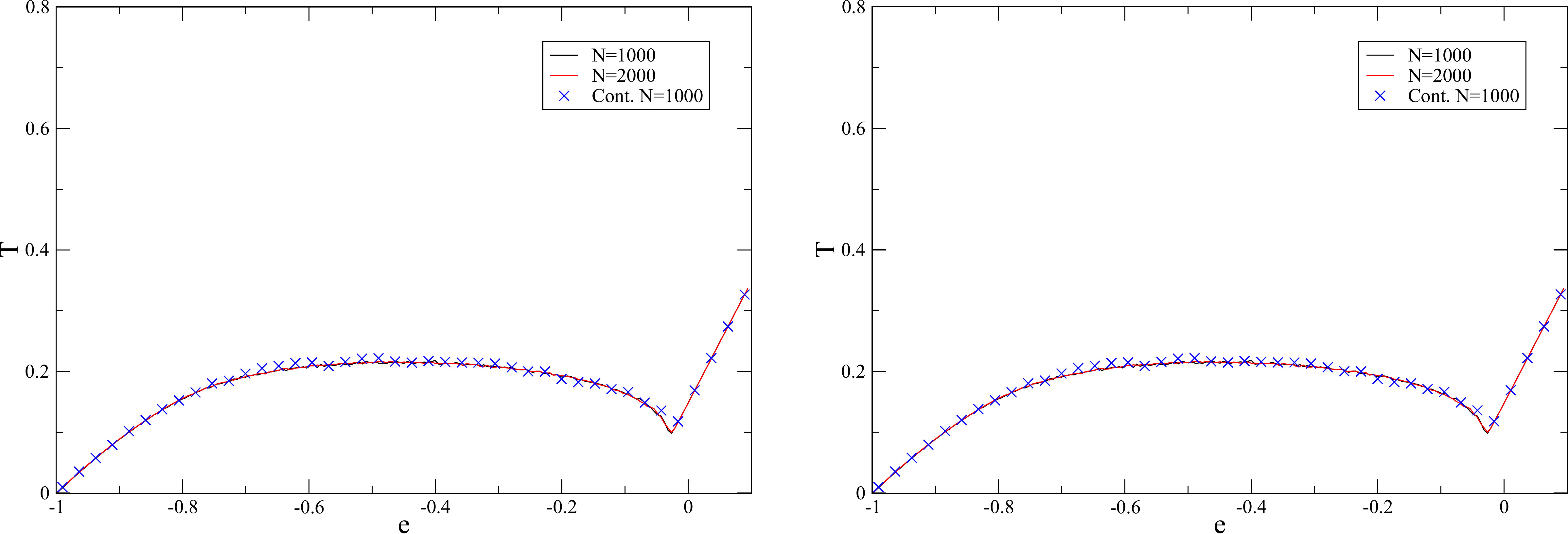}}
\caption{Temperature $T$ as a function of the total energy per particle $e$ for the softening parameter
case with $\epsilon=0.001$ (left panel) and $\epsilon=0.1$ (right panel). In both cases the
curves corresponding to $N=1000$ and $N=2000$ particle coincide. The number of points in the lattice in each case is $N_l=5000$,
the number of Monte Carlo steps per particle is $N_{mc}=5\:000$, and the temperature was averaged over the last $1000$ steps.
Crosses are results obtained for the full (continuous) ring model~\cite{sota} from Monte Carlo simulations with $N=1000$ particles
and $1000$ steps and averaging over the last $100$ steps.}
\label{fig:SPC_caloric}
\end{figure}

\begin{figure}[htbp]
\centering
\scalebox{0.27}{\includegraphics{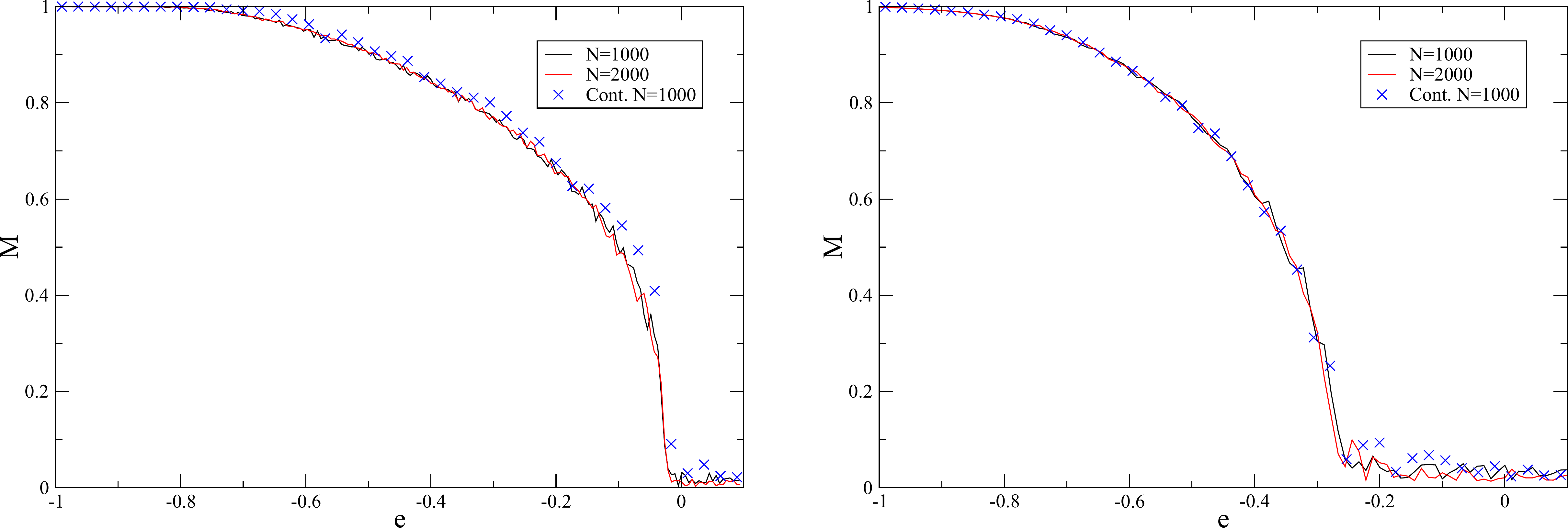}}
\caption{Magnetization $M$ as a function of the total energy per particle $e$ for the softening parameter
case with $\epsilon=0.001$ (left panel) and $\epsilon=0.1$ (right panel). In both cases the
curves corresponding to $N=1000$ and $N=2000$ particle coincide. As in Fig.~\ref{fig:SPC_caloric}
crosses are results for the continuous ring model with $N=1000$.
The simulation parameters are the same as in Fig.~\ref{fig:SPC_caloric}.}
\label{fig:SPC_mag}
\end{figure}

We first consider the self-gravitating ring model with a softening parameter and potential in Eq.~(\ref{sgr_smooth}).
The caloric curves and the magnetization as a function of energy are shown in Figures~\ref{fig:SPC_caloric} and~\ref{fig:SPC_mag}, respectively.
We also show the results obtained from a Monte Carlo simulation for the continuous case,
with a very good agreement and confirming that
a lattice gas description of the model converges rapidly to the continuous case when augmenting the number $N_l$ of lattice points.
Both homogeneous (the straight line part of the graph) and inhomogeneous phases are visible in the figure, with a region of negative heat capacity
for $\epsilon$ sufficiently small. For higher values of the softening parameter (not shown here),
the intermediate energy regime with negative heat capacity disappears and the caloric
curve approaches the one obtained for HMF model for $\epsilon\rightarrow\infty$ (see Eq.~\ref{sgr_li})~\cite{sota,ruffo}.
The fact that all curves coincide for different values of $N$ after rescaling the potential by $|V_{\rm min}|=N/\sqrt{2\epsilon}$
imply that the total energy $E=Ne$ is extensive, which is the reason why the so-called Kac factor $1/N$ was introduced
in the potential in the right-hand side of Eq.~(\ref{hamil_sgl})~\cite{kac}.

\subsection{Hard-core regularization}
\label{sec4b}

\begin{figure}[htbp]
\centering
\includegraphics[width=70mm]{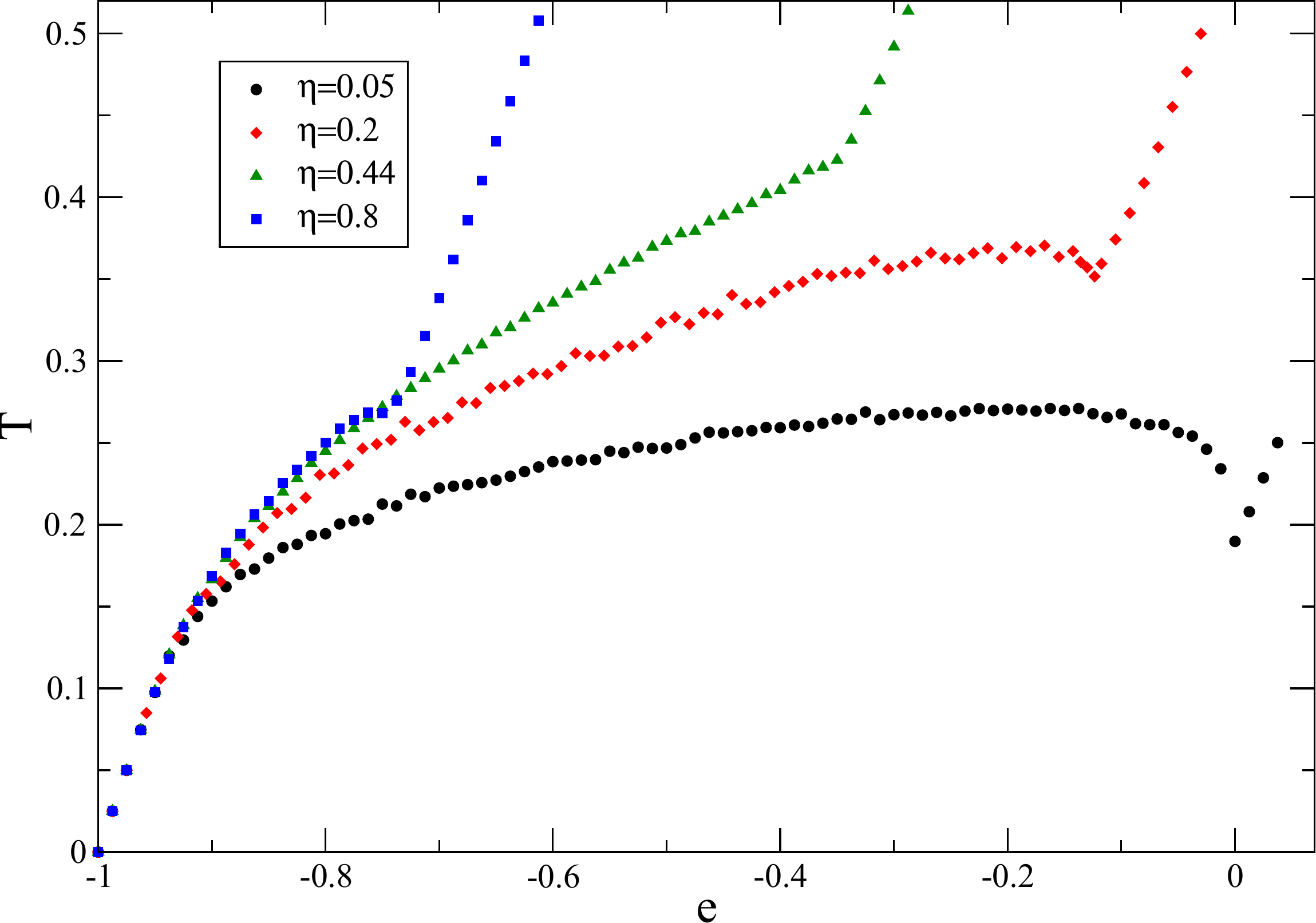}
\caption{Caloric curve for particles with a hard-core in the interparticle potential in Eq.~(\ref{sgl_hard}) for
different packing fractions.}
\label{fig:HPCSGR_c}
\end{figure}

\begin{figure}[htbp]
\centering
\includegraphics[width=80mm]{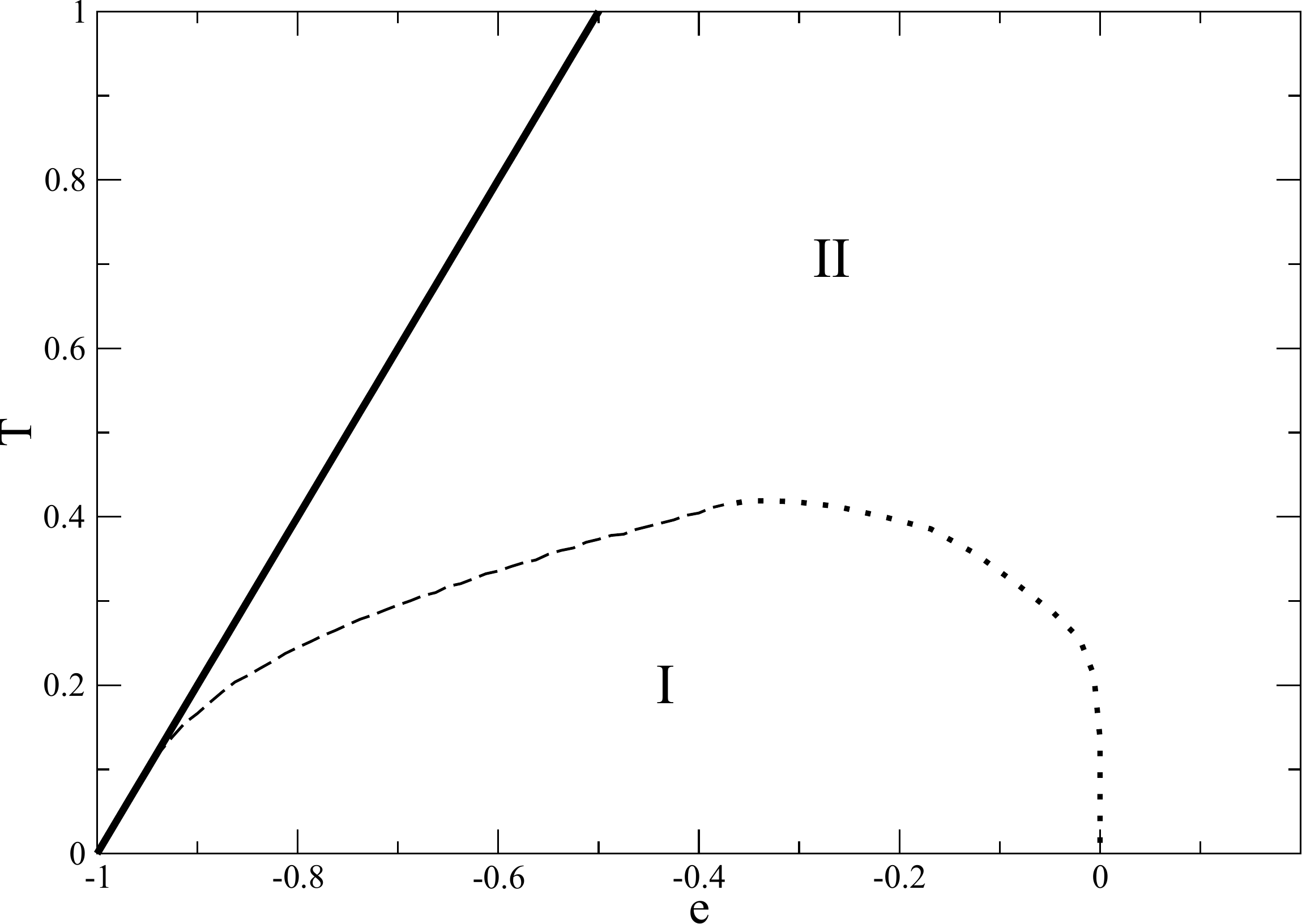}
\caption{Phase diagram for the ring model with a hard-core in the $(e,T)$ plane. In the graphics I and II correspond to the
ferromagnetic (inhomogeneous) and zero magnetization (homogeneous) phases, respectively. The  dotted line represents the
first order and the dashed line the second order phase transition. The thick continuous line is defined by $\eta = 1$
and is the boundary of the physically accessible points in the plane.}
\label{fig:dia_phase2}
\end{figure}

The presence of a hard-core in the inter-particle potential implies that the system has a maximum possible density, given by
$n=N/2\pi$. We associate a finite size to each particle given by the lattice spacing $l$, and define the lattice hard sphere
packing fraction by $\eta=N_l/N=nl$.
Figure~\ref{fig:HPCSGR_c} shows the caloric curves for the potential in Eq.~(\ref{sgl_hard}) for different values of $\eta$.
Here $\eta$ has a similar effect 
as the softening $\epsilon$ in the potential in Eq.~(\ref{sgr_smooth}). As $\eta$ increases, the caloric curves also approach the caloric curve for the
HMF model. For smaller values of $\eta$ a region of negative heat capacity is observed.
A transition from a non-magnetized phase at higher energies
to a ferromagnetic phase at lower energies is also observed, with the critical energy depending on $\eta$.
The phase transition is first order with a temperature jump at the critical temperature for smaller values of $\eta$,
but becomes continuous above a given value of the parameter.
For $\eta\gtrsim0.44$ the negative heat capacity region ceases to exist and the phase transition is continuous. 
The corresponding phase diagram in the $(\eta,e)$ and $(e,T)$ planes are shown in Figs.~\ref{fig:dia_phase1}
and~\ref{fig:dia_phase2}, respectively.

\begin{figure}[htbp]
\centering
\includegraphics[width=70mm]{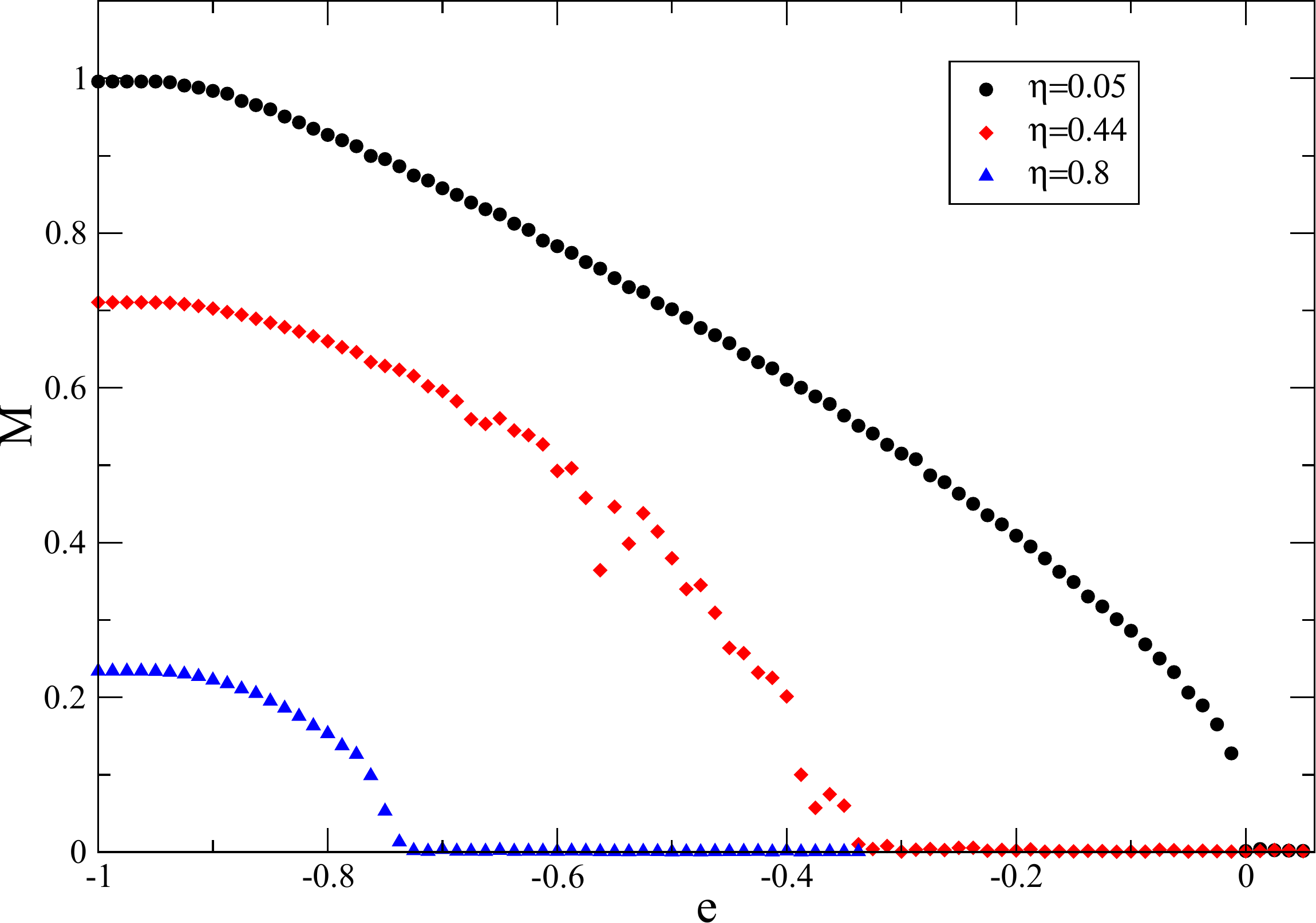}
\caption{Magnetization as a function of energy for a a few values or $\eta$ for the ring model with potential given by Eq.~(\ref{sgl_hard}).}
\label{fig:HPCSGR_mag}
\end{figure}

\begin{figure}[htbp]
\centering
\includegraphics[width=70mm]{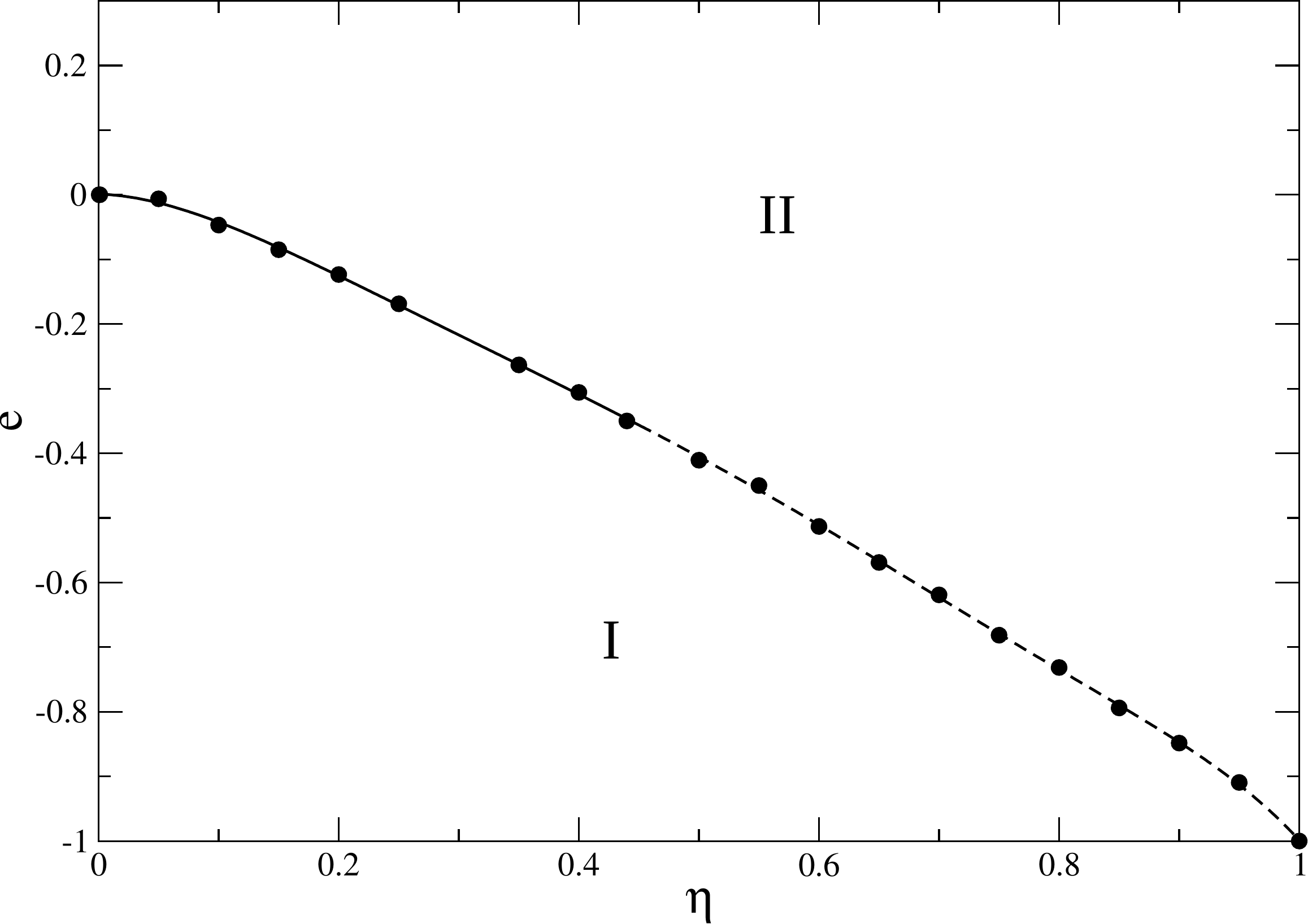}
\caption{Phase diagram for the same system as in Figs.~\ref{fig:HPCSGR_c} and ~\ref{fig:HPCSGR_mag}. Microcanonical Monte Carlo results
are given by the dots in the graph. I stand for the core-halo and II to the homogeneous phase. The solid and dashed lines correspond
to first order and continuous phase transitions, respectively.}
\label{fig:dia_phase1}
\end{figure}

\begin{figure}[htbp]
\begin{center}
\includegraphics[width=70mm]{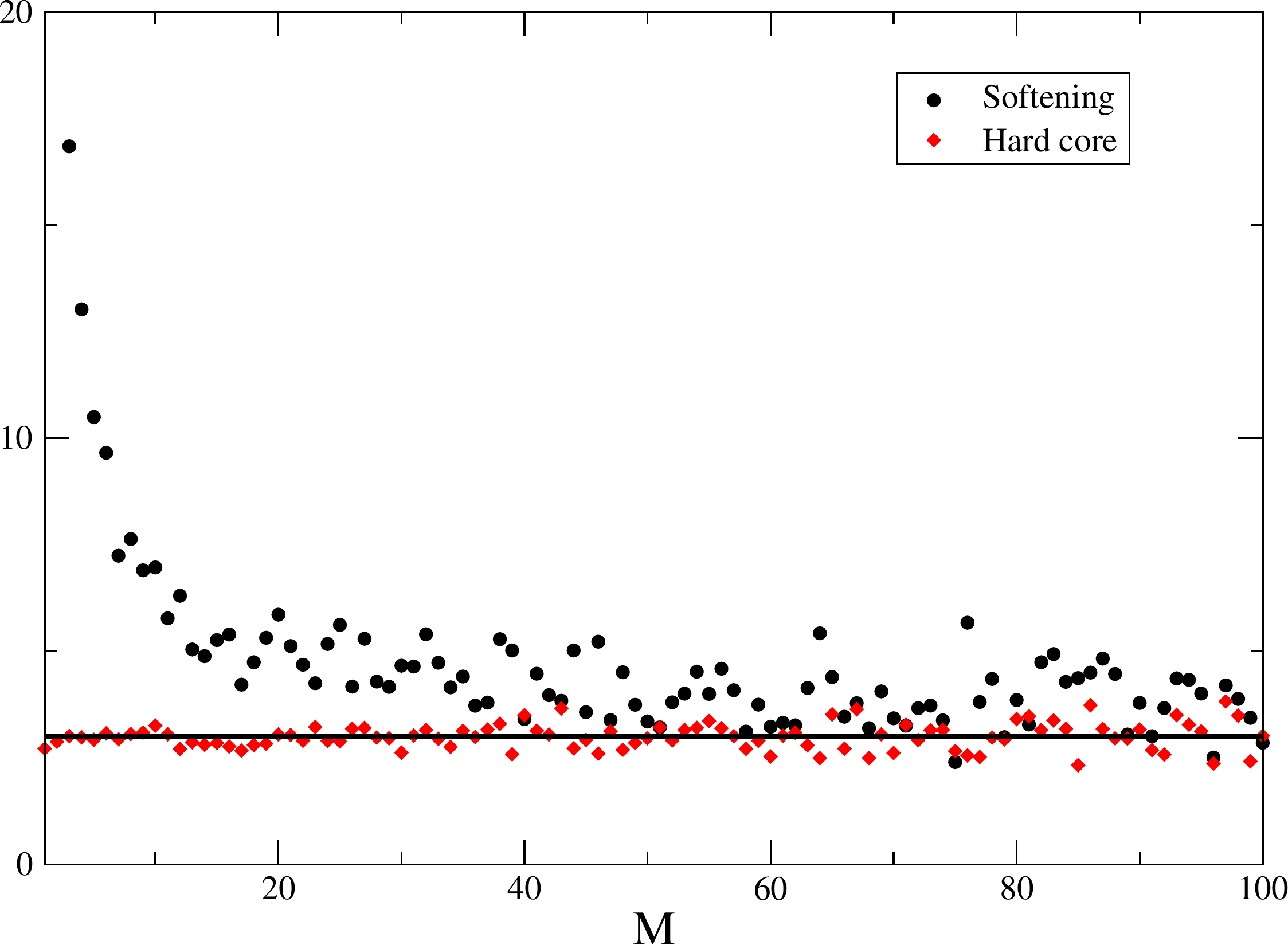}
\end{center}
\caption{Plot of the kurtosis of the sum of $M$ subsequent particle positions in Eq.~(\ref{kurtosistheta}), with
$Y_l=\sum_{k=1}^M\theta_{(k-1)M+l}$ for $l=1,\ldots,{\rm trunc}(N/M)$, ${\rm trunc}(N/M)$ is the closest integer smaller than $M/N$,
as a function of $M$, for a potential with a hard-core and with a softening parameter $\epsilon=10^{-6}$, both cases with $N=10\:000$,
$N_l=20\:000$ and $e=-0.7$. The continuous horizontal line at $K=3$ is drawn for reference.}
\label{interkurt}
\end{figure}

An important point is to consider whether the mean-field description is still valid if the potential has a hard-core. The proof of the exactness of the
mean-field description relies on the regularity of the inter-particle potential~\cite{braun,chavanis,nosbob}, and therefore is not valid for the present case.
The mean-field description is valid if and only if inter-particle correlations are negligible, and this can be tested by a correlation measure used for
the same purpose in Ref.~\cite{nosbob}. Denoting the average of a stochastic variable $Y$ by $\langle Y\rangle$,
its standard deviation and reduced fourth momentum (kurtosis) are defined respectively by:
\begin{equation}
\sigma_Y=\sqrt{\left\langle\left(Y-\langle Y\rangle \right)^2\right\rangle},
\label{standdevtheta}
\end{equation}
\begin{equation}
K_Y=\left\langle\left(\frac{Y-\langle Y\rangle}{\sigma_Y}\right)^4\right\rangle.
\label{kurtosistheta}
\end{equation}
For a variable with a Gaussian distribution  we always have $K_Y=3$. If the position of a given particle is statistically uncorrelated with
the positions of other particles, and from the central limit theorem, the distribution of the sums of particles positions must converge to a Gaussian
distribution. In this way, we partition the set of $N$ particles into subsets
formed by $M$ particles, and then sum the positions of the $N/M$ group of particles
in each subset. The resulting variables must have a Gaussian distribution, or at least close to it, if inter-particles correlations are indeed negligible.
Figure~\ref{interkurt} shows the kurtosis for the distribution of the sum of $M$ subsequent position variables as a function of $M$ for both cases of regularization,
with an energy per particle $e=-0.7$ such that the particles form a compact cluster. For both cases the kurtosis oscillates around the expected value for uncorrelated
variables, indicating that the mean field description is still valid with a hard-core part in the potential, although a further study on the validity of the mean field approach
for long-range interacting potentials with a short-range divergence is still in need~\cite{champion}.

\begin{figure}[htbp]
\begin{center}
\includegraphics[width=70mm]{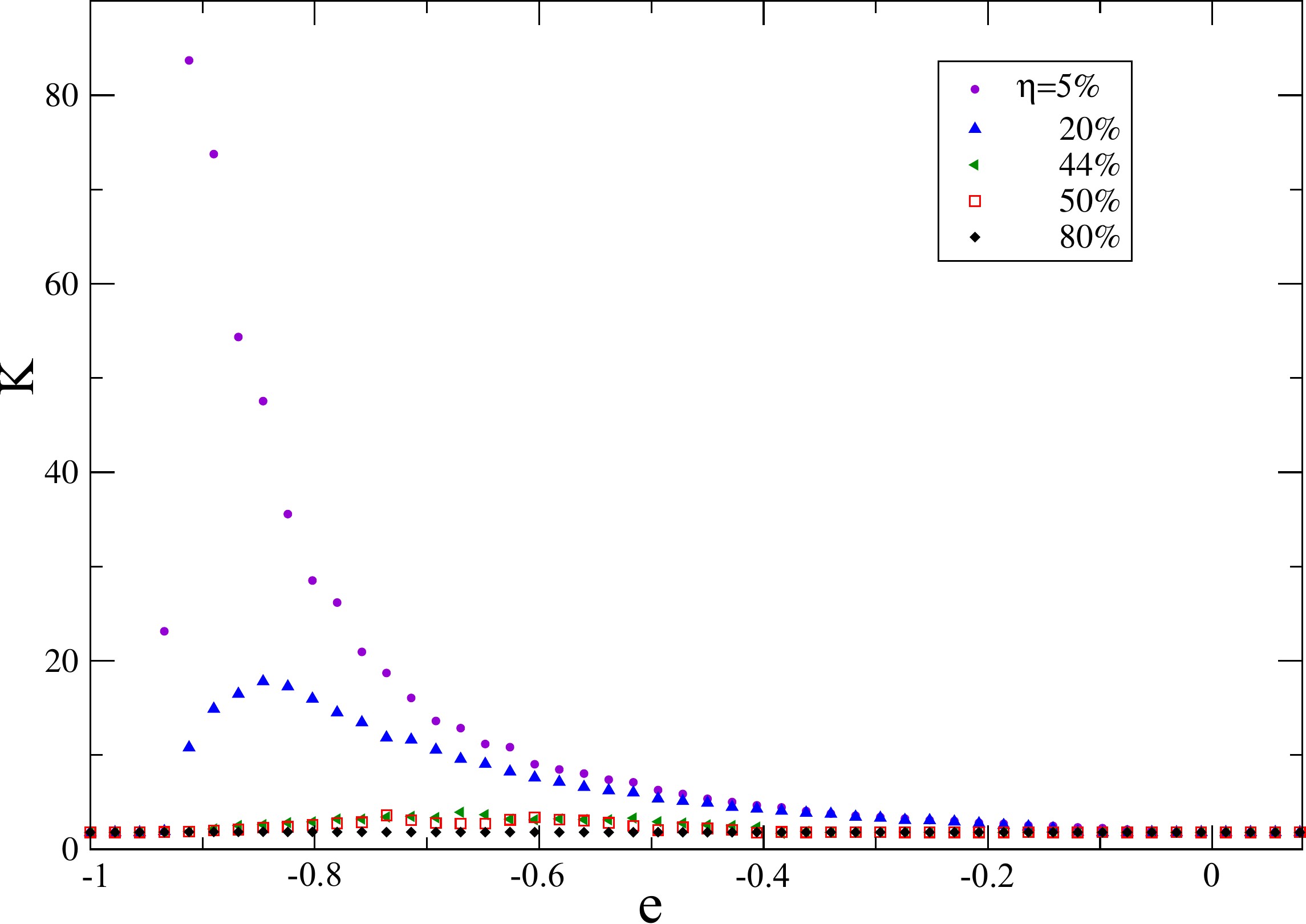}
\end{center}
\caption{Kurtosis $K$ of the spatial distribution of the self-gravitating
ring model with hard-core particles as a function of the energy for different packing fractions,
for $N=5000$ and a variable value for $N_l$.}
\label{figkring}
\end{figure}

The energy landscape method, based on the topological approach to phase transitions,
has been used to predict phase transition in solvable
models with long-range interactions. Base on this approach Nardini and Casetti~\cite{nardini}
proposed a new phase for the self-gravitating
ring model at very low energies (with a phase transition near $e=-1$).
Rocha Filho {\it et al.}~\cite{nosenergland} showed that the self-gravitating ring
model with the softening parameter does not display such low energy phase, nevertheless it displays a marked change in the regime from
a core-only to a core-halo structure. In order to analyze the effect of a hard-core, and whether such a phase is present or not,
we plotted in Fig.~\ref{figkring}
the kurtosis of the self-gravitating ring with hard-core particles as a function of energy.
The kurtosis is very sensitive to minors changes in the distribution function,
and any phase transition should come about as a
discontinuity as a function of the energy~\cite{nosenergland}.
For very low energies ($e<-0.9$), the system is in a
core-only structure, in which the halo is so diffuse that can be considered nonexistent.
For practical purposes all the particles
collapse side by side forming a single core uniform distribution, homogeneous inside a continuous volume $N\ell$ and null otherwise.
For low values of $\eta$, the kurtosis presents a strong maximum in the region $-0.9 \lesssim e <-0.8$.
This corresponds to a change of regime in the distribution
function to a heavy-tail spatial distribution, with a dense core and a diffuse halo.
Despite of this strong peak in the kurtosis, no divergence or discontinuity is present that could be
identified as a phase transition signature. From those results it seems
that a genuine thermodynamic phase transition occurs only for $\eta \rightarrow 0$,
which for $N \neq 0$ corresponds to $\ell \rightarrow 0$ and
the elimination of the hard-core potential and the further collapse of all
particles into a single point due to the short-range divergence of the potential.

\section{$1/r$ linear model}
\label{sec5}

\subsection{Softening parameter regularization}
\label{sec5a}

\begin{figure}[htbp]
\centering
\includegraphics[width=70mm]{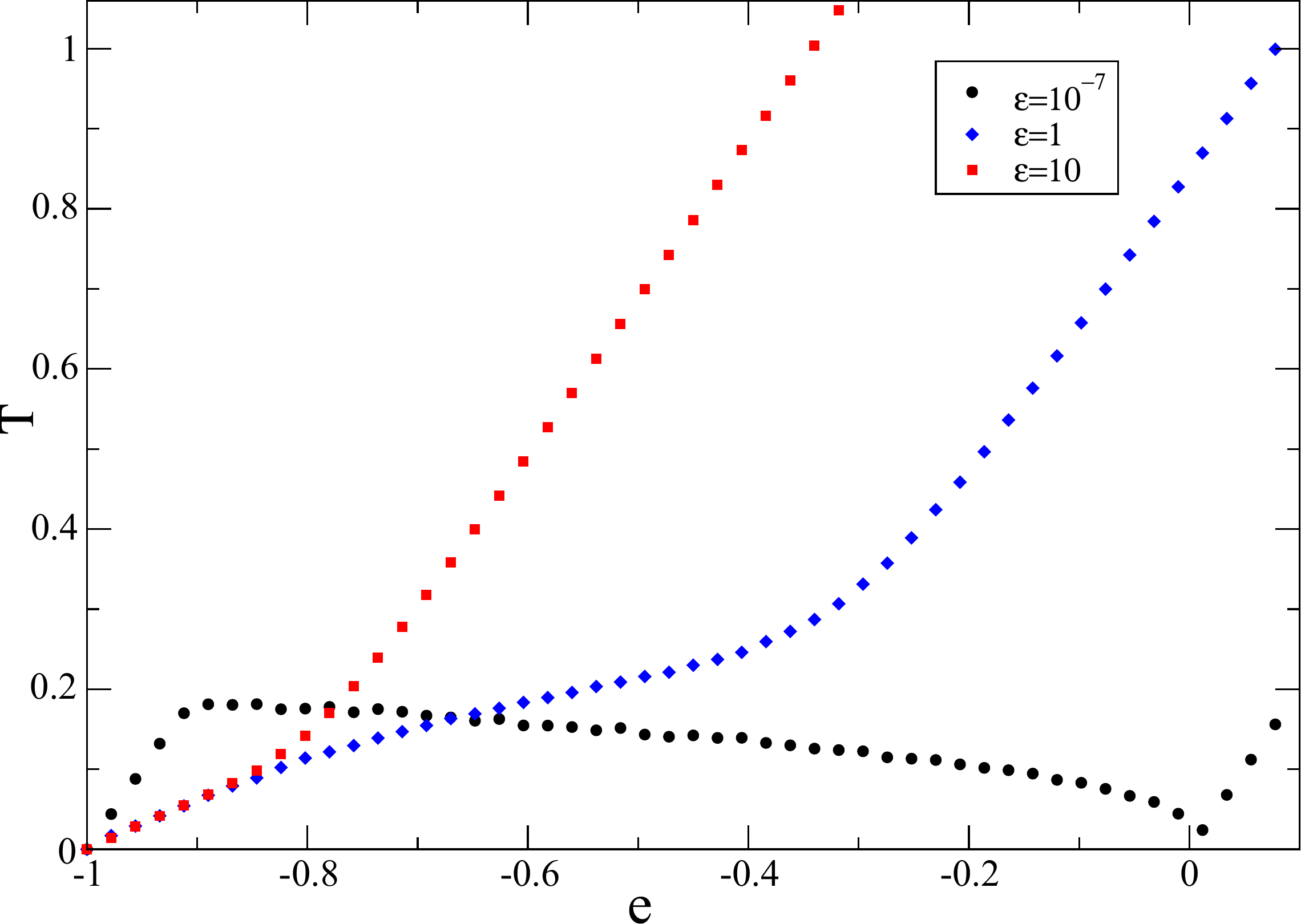}
\caption{Caloric curve for the model with potential in Eq.~(\ref{sgl_smooth})
for different values of $\epsilon$, with $N_l=100\:000$ lattice points and $N=5000$ particles.
For higher values of $\epsilon$ the caloric curve tends to the one for the
one dimensional sheets model. For smaller values of $\epsilon$, it is similar to the one for a three
dimensional self-gravitating systems, with a region with negative heat capacity.}
\label{fig:linesof_c}
\end{figure}

\begin{figure}[htbp]
\centering
\includegraphics[width=70mm]{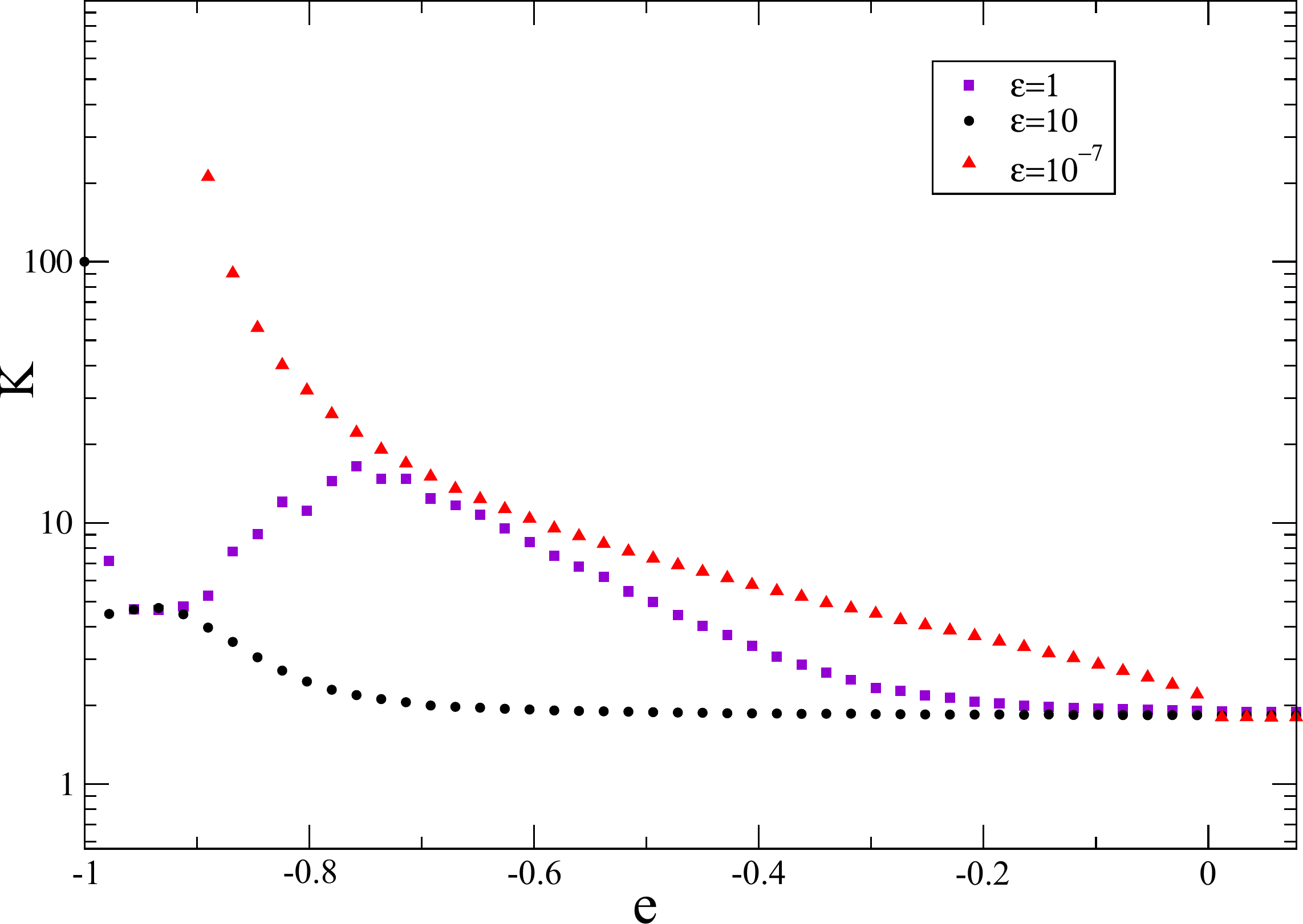}
\caption{Kurtosis of the spatial distribution for the $1/r$ linear model with a
softening parameter as a function of the energy for different packing fractions,
for $N_l=100\:000$ lattice points and $N=5000$ particles.}
\label{fig:linesof_stat_moments}
\end{figure}

Now let us consider a finite system on a straight line with potential given in Eq.~(\ref{sgl_smooth}). This allow us to better grasp the
effect of the topology in the properties of one-dimensional toy models.
Figure~\ref{fig:linesof_c} shows the caloric curves and Fig.~\ref{fig:linesof_stat_moments}
the kurtosis of the spatial distribution as
a function of the energy per particle $e$ for a few values of the softening parameter $\epsilon$.
For smaller values of $\epsilon$ the system has two distinct
phases, a homogeneous phase at higher energies and a clustered phase with a negative heat capacity region at lower energies,
with a first order phase transition. The high energy phase is characterized by positive and constant heat capacity and a
value $K=1.8$ for the kurtosis, implying a homogeneous phase. 
The low energy phase has two different regimes although not separated by a phase transition:
the lower energy regimes with a core-only structure and positive heat capacity region stabilized by the softening parameter,
and a regime in a region of intermediate energy with a core-halo structure with a region with negative heat capacity~\cite{nosenergland}.
These two configurations are separated by a peak of higher values of the kurtosis, but without any visible divergence.
The height of this peak becomes greater for smaller $\epsilon$, such that
in the limit $\epsilon \rightarrow 0$ the change in the regime of the distribution function becomes a genuine thermodynamic phase transition.

For higher values of $\epsilon$, the heat capacity is always positive and the caloric curve tends to the one for the one-dimensional sheets model,
with only one phase for all energy values and with the system going from a clustered to a homogeneous distribution regimes without any
phase transition, as expected from the discussion in Sec.~\ref{sec2}.

\subsection{Hard-core regularization}
\label{sec5b}

For the potential given in Eq.~(\ref{sgl_hard}) we also define the packing fraction parameter as:
\begin{equation}
\label{occupline}
\eta\equiv\frac{N}{N_l}=\frac{N \ell}{L},
\end{equation}
with $\ell$ the distance between two lattice points taken as the size of the hard-core (the size of the particles) and $L=2 \pi$.

\begin{figure}[htbp]
\centering
\includegraphics[width=70mm]{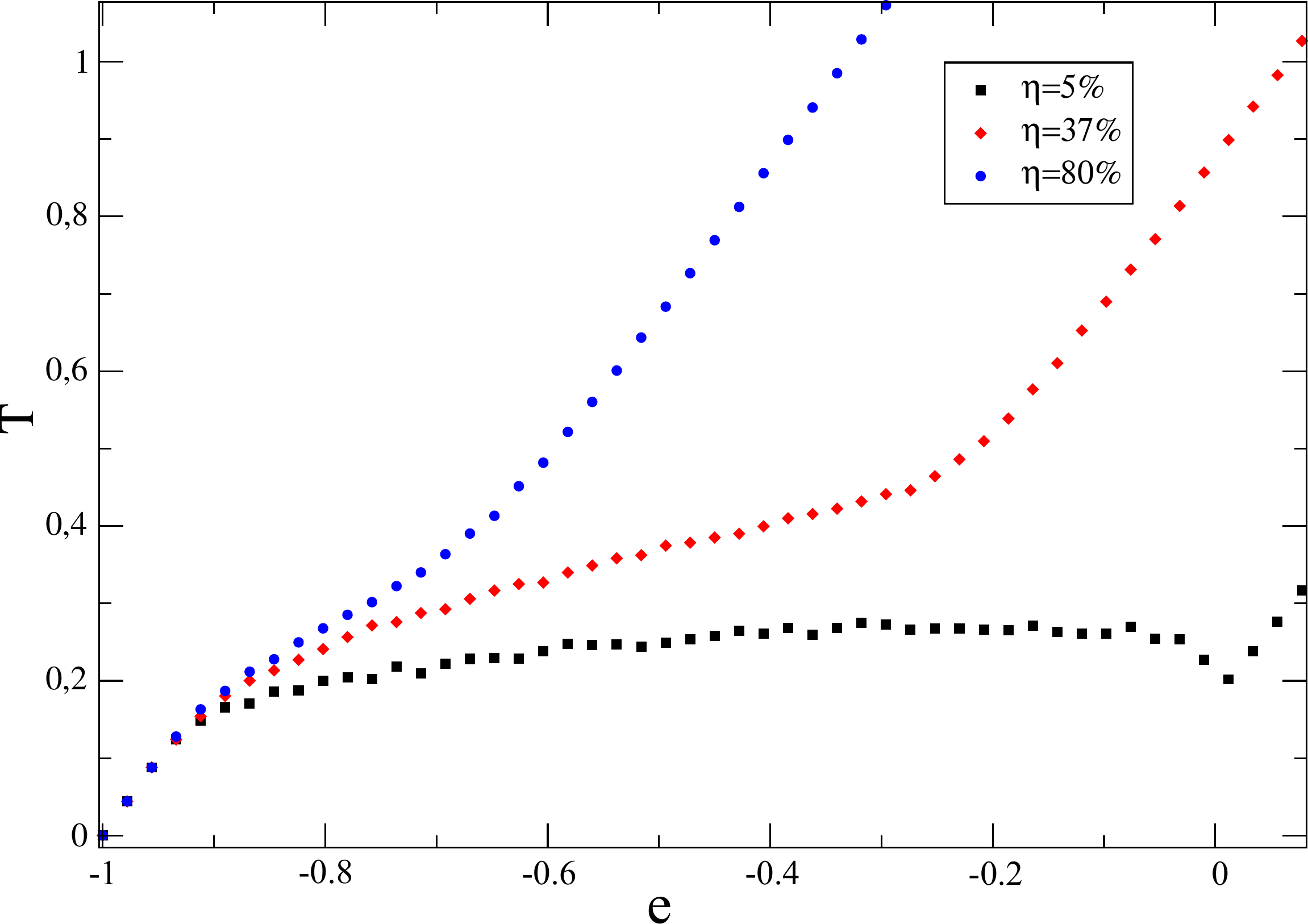}
\caption{Caloric curve for the linear model model with a hard-core for different packing fractions $\eta$
and number of particles $N=1000$.}
\label{fig:linehard_c}
\end{figure}

\begin{figure}[htbp]
\centering
\includegraphics[width=70mm]{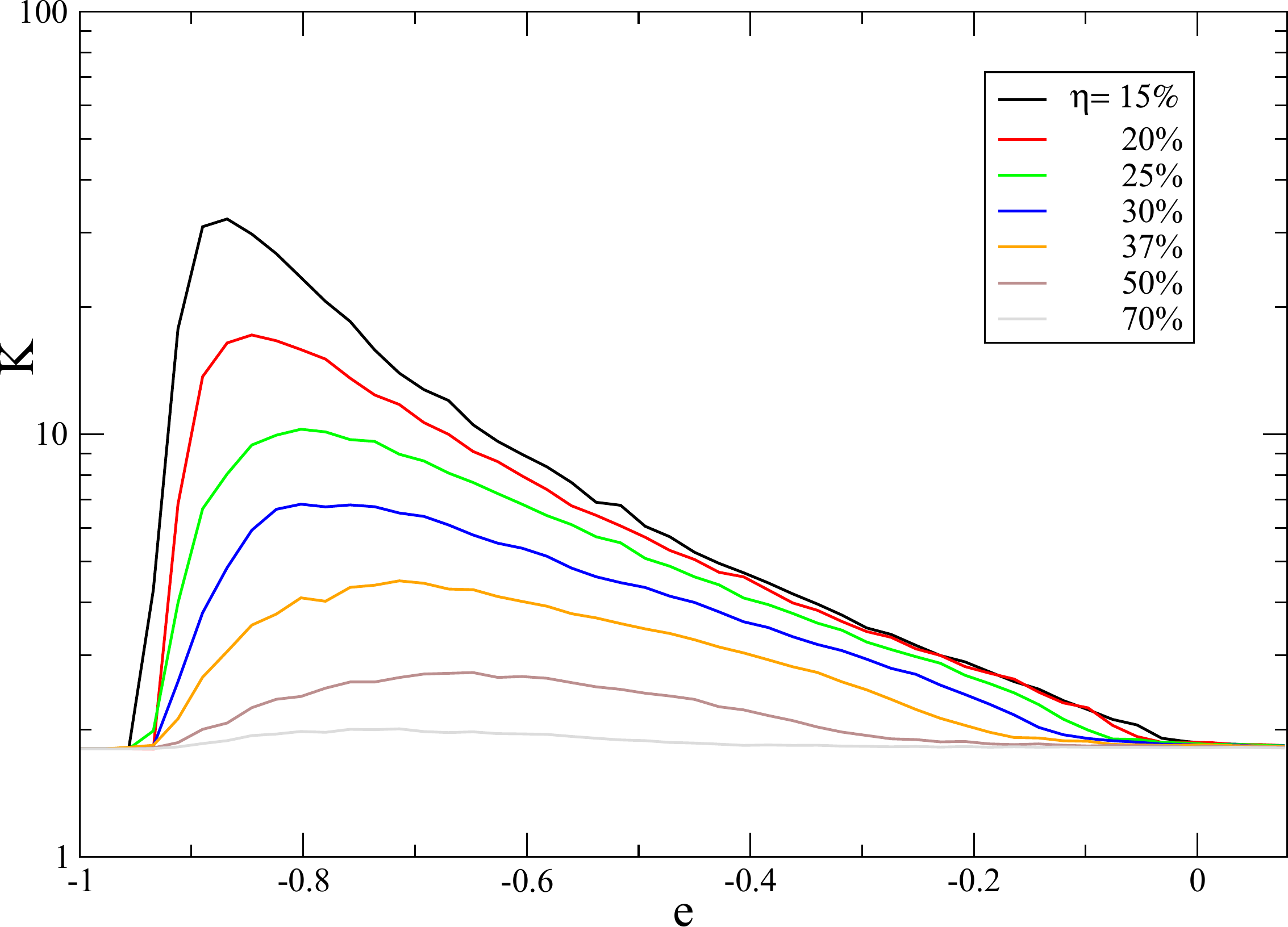}
\caption{Kurtosis of the spatial distribution for the $1/r$ linear
model with hard-core particles as a function of the energy for different packing fractions, for $N=5000$
particles.}
\label{fig:linehard_stat_moments}
\end{figure}

Figure~\ref{fig:linehard_c} shows the caloric curves and Fig.~\ref{fig:linehard_stat_moments}
the kurtosis for the spatial distribution as a function of
the energy, for some values of $\eta$. Similarly to the ring model case, different values of $N$ and $N_l$ but
with same value of $\eta$ have the same caloric curve, provided the energy is rescaled by the absolute value of the potential as explained above.

Similarly to the ring model with hard-core particles, the parameter $\eta$ has a similar effect as the softening parameter $\epsilon$.
For lower values of $\eta$ ($\eta<30\%$) the system has
two phases: a high energy homogeneous
phase with constant positive heat capacity and a low energy clustered phase with a region of negative heat capacity in an
intermediate energy range, with a first order phase transition with critical energy depending on $\eta$. Here also the clustered phase
has two spatial distribution regimes: a core-only structure regime stabilized by the hard-core
potential and positive heat capacity for very low energies, while for intermediate energies
a core-halo structure is observed with a negative heat capacity region in the energy. The marked change between these two regimes shows 
the formation of the strong maximum of the kurtosis that for $\eta \rightarrow 0$ or, equivalently,  for $\ell \rightarrow 0$
the peak in the kurtosis diverges and
the change between the two regimes becomes a genuine thermodynamic phase transition.

\begin{figure}[htbp]
\centering
\includegraphics[width=130mm]{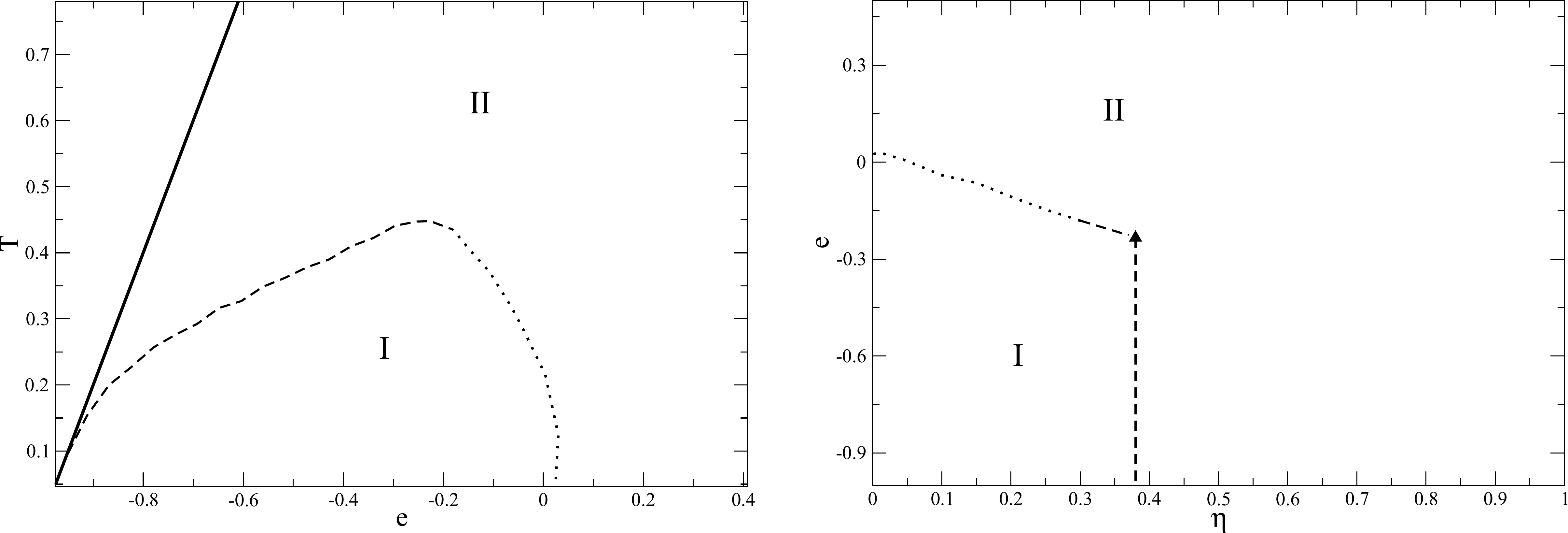}
\caption{Phase diagram for the $1/r$ linear model with hard-core particles in the $(e,T)$ plane (left panel) and in the $(\eta,e)$ plane (right panel).
In the graphics I and II correspond to the
clustered (inhomogeneous) and homogeneous phases, respectively. The dotted line represents the
first order and the dashed line the second order phase transition. The continuous line in the
left panel is defined by $\eta = 1$ and it is the boundary of the physically accessible points in the $(e,T)$
plane and the triangle point in the right panel is the critical packing fraction $\eta_{c}$ above which the system
tends to the behavior of the one-dimensional sheets model with only one phase to all energy values.}
\label{fig:diagrama_line_hard}
\end{figure}

For values of $\eta$ larger than some critical value $\eta_{c}$, with $\eta_{c}\approx37\%$, the behavior is similar to the one-dimensional
sheets model and, as said previously, it has only one phase and the system goes continuously from the clustered to the homogeneous regime.
For intermediate packing fractions $30\%\lesssim\eta< \eta_{c}$, the system has a transient behavior between the  one characterizing a one-dimensional sheet
model and a reminiscent behavior of a true self-gravitating system, with two phases:
a high energy homogeneous phase and a clustered phase which a positive heat capacity,
and a second order phase transition.
The corresponding phase diagram is shown in Fig.~\ref{fig:diagrama_line_hard}.

\section{Extensivity of the energy}
\label{sec6}

One of the main consequences of the long-range nature of the interactions is the non-additivity of the entropy, which has profound implications
on the thermodynamic equilibrium properties of such systems, as already discussed in the introduction. Although non-additive, the extensivity
of the entropy is usually restored by the introduction of the Kac factor $1/N$ in the potential energy~\cite{kac}. Since the potential energy has
$N^2$ terms (for large $N$), each contributing with the same weight to the average potential energy and thus resulting
in a $N^2$ factor, the introduction of the Kac factor implies
a potential energy proportional to $N$. Since the kinetic energy has $N$ terms, the total energy is consequently extensive,
as the entropy. This is an important to have a well defined thermodynamic limit,
also called the Vlasov limit for long-range systems. It is also used to
demonstrate the validity of the Vlasov equation to describe the non-equilibrium dynamics of such systems~\cite{nosbob,braun}.
This is valid for regular interparticle potentials, i.~e.\ continuous and with no divergence.
This is verified for the models in Sections~\ref{sec4a} and~\ref{sec5a}
with a softening parameter regularization. Since all particles can occupy the same lattice point as the lowest energy state,
$|V_{min}|$ is proportional to $N$ and independent of $N_l$,
and owing to the fact that all intensive thermodynamics properties (e.~g.\ the temperature and energy per particle) coincide,
up to fluctuations, for all values of $N$ (when the potential is rescaled by $|V_{min}|$), we conclude
that the internal energy is extensive.

\begin{figure}[htbp]
\centering
\includegraphics[width=140mm]{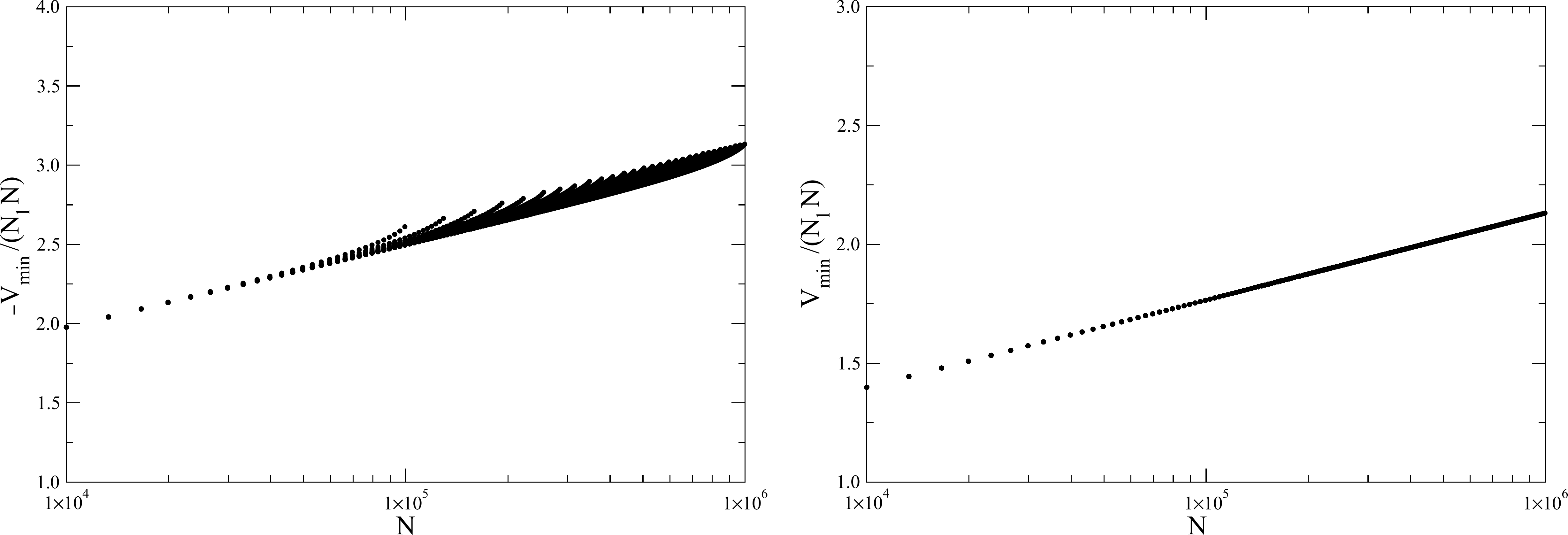}
\caption{Left Panel: Plot of $|V_{min}|/N_lN$, for the self-gravitating ring model with a hard-core in the potential,
from maximal particle packing as a function of $N$ for values of $N$ from $10^4$ to $10^6$
and number of lattice points $N_l$ from $10^5$ to $10^6$.
Right Panel: Same as the left panel but for the linear $1/r$ model.}
\label{vminring}
\end{figure}

Now, if for instance, the potential has a hard-core, the simplicity of the Kac prescription is lost.
To illustrate this point, we show in Fig.~\ref{vminring}
the plot if $|V_{min}|/N_lN$ as a function of $N$ for the ring and the $1/r$ finite model, both with a hard-core in the potential.
The fact that the plots are not horizontal lines implies then that the usual Kac prescription is not sufficient to have
an extensive energy.
In the case of the $1/r$ potential with hard-core particles, the minimum potential can be computed analytically
with all the particles side by side in the maximum clustering possible.
In this configuration one can rewrite the double sum in Eq.~(\ref{hamil_sgl}) as a single sum over
the contribution of all the $i$-th neighbors, i. e., there are $N-i$ particles in a distance $i \ell$ from each other:
\begin{equation}
\label{vmin_1}
V_{min}= - \sum \limits_{i=1}^{N-1} \frac{N-i}{i \ell}=\frac{1}{\ell} \left[ N-1 - N\; \sum \limits_{i=1}^{N-1} \frac{1}{i} \right],
\end{equation}
the sum in the last term of Eq.(\ref{vmin_1}) is the partial harmonic sum, that for $N \gg 1$ yields:
\begin{equation}
\label{vmin_2}
\vert V_{min} \vert\propto \left(\frac{N^2}{\eta L}\right) \ln(N).
\end{equation}
Therefore for the potential to be extensive (proportional to $N$) the Kac factor must be modified to $1/(N\ln N)$.
This is another result of our approach.
For the ring model it is not yet clear what this modification should be performed.

\section{Concluding Remarks}
\label{sec7}

The hard-core part of the potential introduces the effect of a minimum possible volume for the system, which is equivalent
to establish a particle size. The packing fraction parameter $\eta$ plays a role similar to
a softening parameter used to regularize the zero distance divergence of the attractive potential. For lower values of $\eta$
the caloric curve and phase diagram is similar to 
the three dimensional gravitational system, with a negative heat capacity.
For higher values of $\eta$ the same curve for the ring model with
a hard-core in the potential approaches the caloric curve of the HMF, while for the $1/r$ model it approaches the curve
for the sheets model in the same limit.

We presented the phase diagram for both models. In the case of a potential with a hard-core, we noticed the existence of a critical value $\eta_{c}$
of the packing fraction parameter such that for $\eta<\eta_c$ the system has a first order 
phase transition. For the ring model with a hard-core this critical value is $\eta_{c}=0.44$ and $\eta_{c}=0.35$ for the linear model.
The packing necessary thus depends on the topology of the system, with the potential always proportional to the inverse of the distance.

We also discussed that the Kac prescription must be modified if the interparticle potential has a hard-core, which was explicitly obtained
for the $1/r$ model. Further research on this point is the subject of ongoing research.

\section{Acknowledgments}

J.~M.~Maciel is supported by CAPES.  M.~A.~Amato, T.~M.~Rocha Filho and A.~Figueiredo are partially financed by CNPq (Brazilian government agencies).

\end{document}